\documentclass[prl,twocolumn,showpacs,groupedaddress]{revtex4}
\def\ba{\begin{eqnarray}}
\def\ea{\end{eqnarray}}

\usepackage{graphicx}
\usepackage{dcolumn}
\usepackage{bm}
\usepackage{epsf}

\begin{document}

\title{Dynamical Determination of the Innermost Stable Circular Orbit
of Binary Neutron Stars}

\author{Pedro Marronetti}
\altaffiliation{Fortner Fellow}
\affiliation{Department of Physics, University of Illinois at 
Urbana-Champaign, Urbana, IL 61801}

\author{Matthew D. Duez}
\affiliation{Department of Physics, University of Illinois at 
Urbana-Champaign, Urbana, IL 61801}

\author{Stuart L. Shapiro}
\altaffiliation{Department of Astronomy \& NCSA, University of 
Illinois at Urbana-Champaign, Urbana, IL 61801}
\affiliation{Department of Physics, University of Illinois at 
Urbana-Champaign, Urbana, IL 61801}

\author{Thomas W. Baumgarte}
\altaffiliation{Department of Physics, University of Illinois at 
Urbana-Champaign, Urbana, IL 61801}
\affiliation{Department of Physics and Astronomy, Bowdoin College, 
Brunswick, ME 04011}

\begin{abstract}
We determine the innermost stable circular orbit (ISCO) of binary
neutron stars (BNSs) by performing dynamical simulations in full
general relativity.  Evolving quasiequilibrium 
(QE) binaries that begin at different 
separations, we bracket the location of the ISCO by
distinguishing stable circular orbits from unstable plunges.  We study
$\Gamma=2$ polytropes of varying compactions in both corotational and
irrotational equal-mass binaries.  For corotational binaries we find an
ISCO orbital angular frequency somewhat smaller than that 
determined by applying
turning-point methods to QE initial data.  For the
irrotational binaries the initial data sequences
terminate before reaching a turning point, but we find that 
the ISCO frequency is reached prior to the termination point.
Our findings suggest that the ISCO frequency varies with compaction but 
does not depend strongly
on the stellar spin. Since the observed gravitational wave signal undergoes
a transition from a nearly periodic ``chirp'' to a burst at roughly twice
the ISCO frequency, the measurement of its value by laser interferometers
(e.g LIGO) will be important for determining 
some of the physical properties of the underlying stars.

\end{abstract}


\maketitle



The emission of gravitational radiation drives the slow inspiral of
neutron star and black hole binaries towards their coalescence and
merger.  Fully general-relativistic numerical simulations are required
for the accurate description of the late inspiral and plunge epochs of
the binary evolution (see, e.g., \cite{Baumgarte:2002jm} for a recent
review).  While complete orbits of binary black holes have not been
numerically simulated yet, simulations of binary neutron star (BNS)
mergers are now becoming sufficiently mature to provide results of
astrophysical interest (e.g.~\cite{Shibata_Uryu}).

One piece of information of great astrophysical interest is the
location of the innermost stable circular orbit (ISCO).  The evolution
of a binary system occurs in three distinct phases \cite{gradual}: (1) A slow,
adiabatic inspiral phase that is driven by gravitational radiation reaction forces 
and can be approximated as a sequence of quasi-circular orbits; 
(2) a brief transition phase, where the inward radial motion increases and the 
orbital motion changes from slow inspiral to rapid plunge; (3) a 
plunge phase, terminating in the merger of the objects. The ISCO
resides within the transition region; its identification is complicated by
the fact that it is not arbitrarily sharp and cannot be localized precisely.  
The gravitational wave quasi-periodic ``chirp'' signal of the 
slow binary inspiral ends at about the twice the orbital angular frequency of 
the ISCO, where it changes its form to a wave signal characteristic of a burst 
(compare~\cite{DBSSU01}).

Within the framework of Newtonian and post-Newtonian gravity, the ISCO
has been determined by different methods (see, e.g., the
reviews \cite{Rasio:1999ku,Baumgarte:2001ab} and references therein).
Much less is known for fully relativistic binaries.  For {\em
corotational} binaries, a turning point on a curve of the binding energy 
vs. separation for 
quasiequilibrium (QE) models along a sequence of constant rest mass marks
the onset of {\em secular} instability \cite{Friedman:1988,BCSST98}.
In the following we will refer to this point as the QE-ISCO.  No such
theorem exists for {\em irrotational} binaries or for the onset of 
{\em dynamical} instability \cite{footnote0}.  Locating the ISCO 
dynamical instability therefore
requires dynamical evolution simulations of the full set of Einstein's
equations for the gravitational field, coupled to relativistic 
hydrodynamics in the case of BNSs.

In this paper we present the first attempt to dynamically locate this
ISCO.  We identify BNS configurations that correspond to stable and
unstable circular orbits by evolving binary initial data sets for different
separations.  The objective is to bracket the location of the ISCO by
distinguishing configurations that can maintain quasi-circular motion
for more than one orbital period and systems that plunge and coalesce
in a fraction of that time.


We adopt the QE initial data presented by Marronetti and
Shapiro \cite{Marronetti:2003gk}, describing two identical neutron
stars in quasi-circular orbit.  These data have been constructed using
the conformal thin-sandwich (CTS) decomposition of the constraint
equations, together with maximal slicing and spatial conformal
flatness.  The formalism introduced in \cite{Marronetti:2003gk} allows
for a free specification of the spin of the stars 
in an approximate fashion.  In this paper we
consider corotational binaries as well as ``irrotational'' binaries
\cite{footnote1} with zero (equatorial) fluid circulation, which are
believed to be more realistic astrophysically \cite{Kochanek:1992}.
Sequences of corotating binaries feature a minimum in the binding 
energy at the QE-ISCO.  Irrotational binaries, however, typically
terminate before reaching this minimum (see Figs.~7 and 8 in 
\cite{Marronetti:2003gk}; also \cite{USE00,GGTMB01} as well as
\cite{Baumgarte:2002jm} for more references).

We adopt a $\Gamma = 2$ polytropic equation of state (EOS) for
which the maximum rest mass
(gravitational mass) of a star in isolation in nondimensional units
\cite{footnote2} is $m_0 = 0.180$ ($m = 0.164$) with a compaction
ratio of $(m/R)_\infty = 0.216$ .  We study models that have two
different compaction ratios in isolation: a moderate value
$(m/R)_{\infty} = 0.142$ (both corotational and irrotational binaries;
cases A and B) and a high value $(m/R)_{\infty} = 0.195$ (only
irrotational binaries; case C).  These compaction ratios correspond to
individual stars with rest masses $m_0=0.1469$ and $m_0=0.1767$,
respectively. The fully general relativistic hydrodynamical code employed for this
study has been introduced in Duez {\it et al.} \cite{Duez:2002bn}.  We
evolve the gravitational fields using the BSSN formalism \cite{BSSN}
with a Courant factor of 0.46.  We approximate maximal slicing with a
``K-driver'' and use a ``Gamma-driver'' shift condition that keeps
$\bar \Gamma^i \equiv \bar \gamma^{lm} \bar \Gamma^i_{lm}$
approximately constant. The simulations were performed on uniform
Cartesian grids in a reference frame that rotates with the binary,
which improves conservation of angular momentum
\cite{Swesty:1999ke,Duez:2002bn} and reduces the
spurious eccentricities of the stable orbits. In all cases presented here 
the spatial volume covered by the grids is completely enclosed by the
light cylinder (defined by coordinate radius $R_L = 1/\Omega_{\rm
orb}$).  A more detailed description of the evolution of the
gravitational and hydrodynamical fields, the boundary conditions and
their numerical implementation can be found in Duez {\it et al.}
\cite{Duez:2002bn}.



\begin{figure}
\epsfxsize=3.0in
\begin{center}
\leavevmode \epsffile{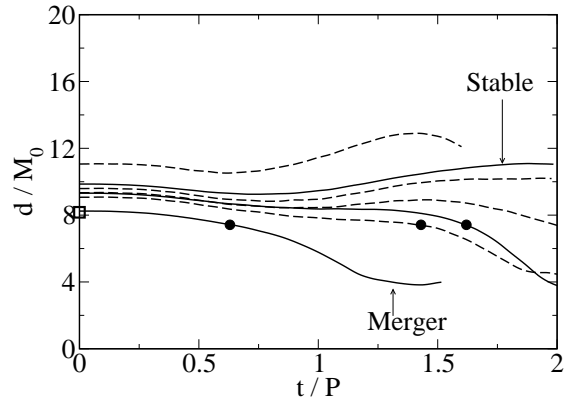}
\end{center}
\caption{ Coordinate
separation vs. time for sequences A.  The separation $d$ is the
coordinate distance between points of maximum rest mass density and is
given in units of the total rest mass of the binary $M_0$. The curves
show runs with different grid sizes and resolutions which are detailed
in Table \ref{Table_Grids}. The filled circles mark the time of surface
contact for the merger orbits. The empty square on the $y$ axis marks 
the QE estimation of the ISCO separation.}
\label{CR_0.14_a_1.0_Sep}
\end{figure}
\begin{figure}
\epsfxsize=3.0in
\begin{center}
\leavevmode \epsffile{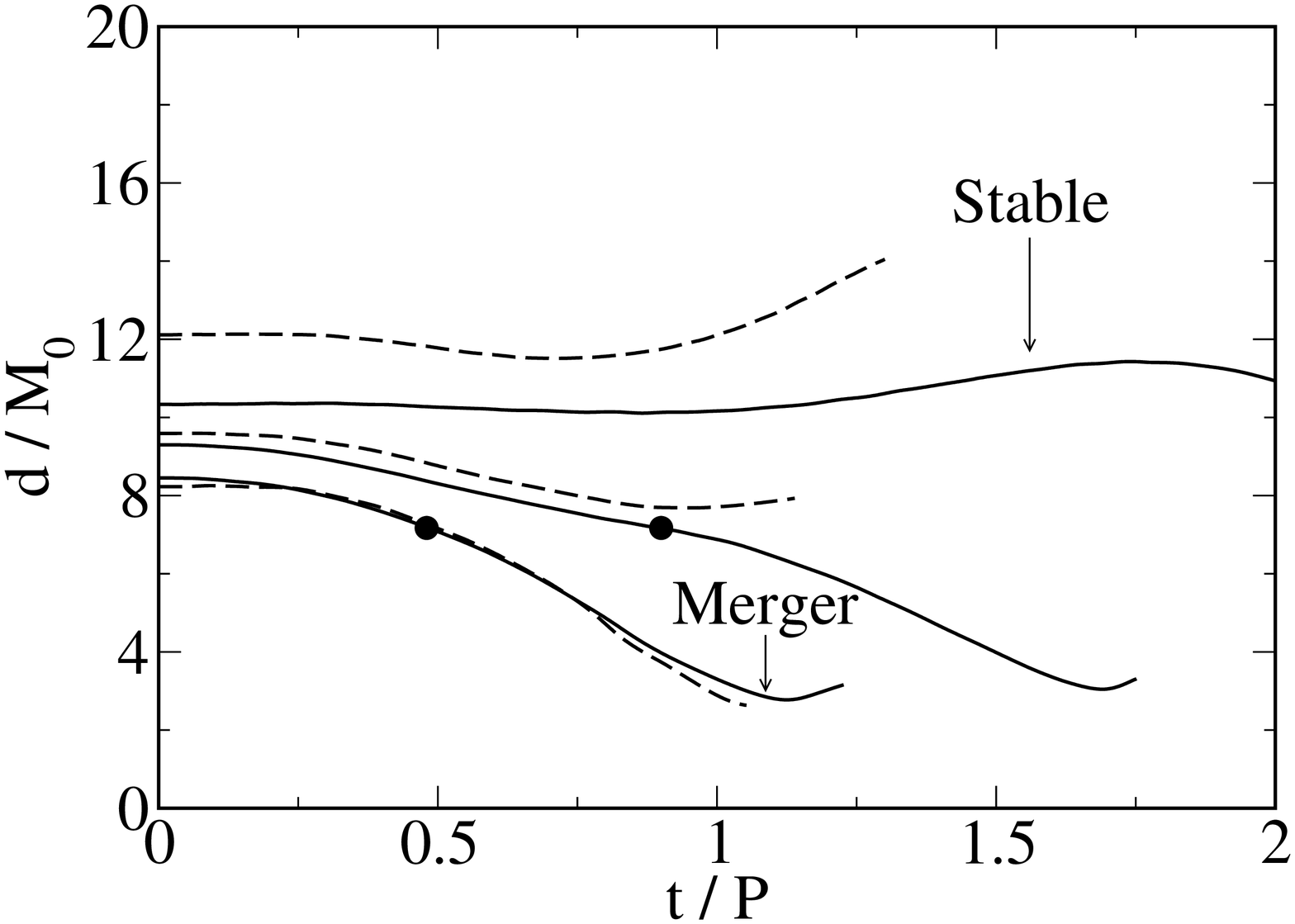}
\end{center}
\caption{Coordinate separation vs. time for sequences B.}
\label{CR_0.14_c_0.0_Sep}
\end{figure}
\begin{figure}
\epsfxsize=3.0in
\begin{center}
\leavevmode \epsffile{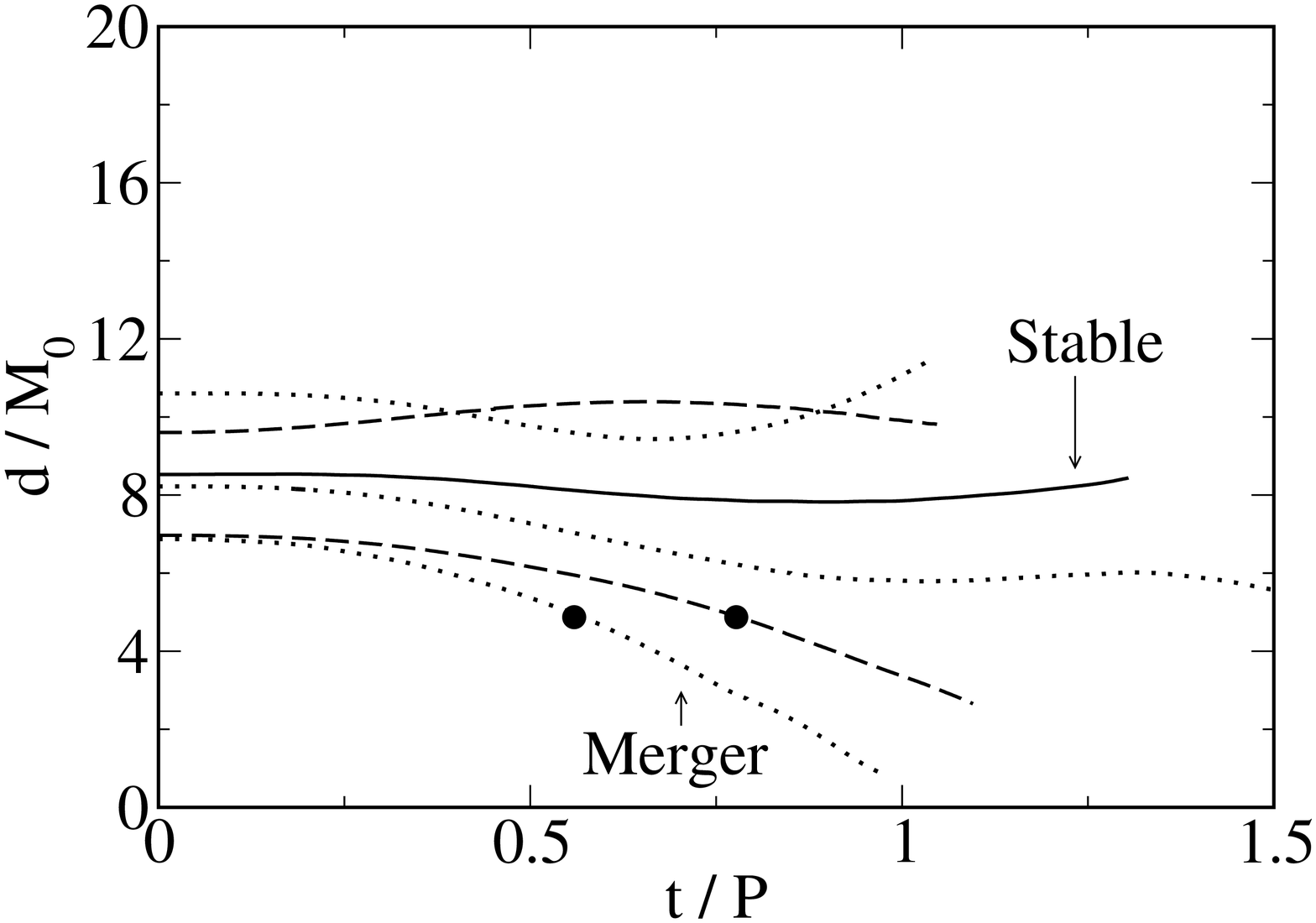}
\end{center}
\caption{Coordinate separation vs. time for sequences C.}
\label{CR_0.19_c_0.0_Sep}
\end{figure}

{\bf Results~~}Figures \ref{CR_0.14_a_1.0_Sep}, \ref{CR_0.14_c_0.0_Sep}, and
\ref{CR_0.19_c_0.0_Sep} show the evolution of the coordinate
separation $d$ between maximum baryonic density points in each star
for cases A, B and C \cite{footnote3}.  The filled circles mark the
points of surface contact for those runs that result in a merger.
In each plot we present results for different grid resolutions and
bounding box sizes, which are listed in Table \ref{Table_Grids}.  We
use the highest quality results to bracket the ISCO; these are labeled
as {\it Stable} and {\it Merger} in the plots.  Results obtained on
smaller computational grids agree with these brackets fairly well.
Note that all three merger cases experience surface contact (and the
related mass interchange) well after the start of the inspiral plunge.
\begin{table}
\begin{center}
\begin{tabular}{cccccccc}
Case & Fig. & Curves & Res. & $B/M_0$ & Grid points & Res. & Box\\
\tableline
A & \ref{CR_0.14_a_1.0_Sep} & dashed & 20 & 18.7 & $64^2 \times 128$  
	& Low  & Large \\
A & \ref{CR_0.14_a_1.0_Sep} & solid  & 40 & 18.7 & $128^2 \times 256$ 
	& High & Large \\
B & \ref{CR_0.14_c_0.0_Sep} & dashed & 20 & 18.7 & $64^2 \times 128$  
	& Low  & Large \\
B & \ref{CR_0.14_c_0.0_Sep} & solid  & 40 & 18.7 & $128^2 \times 256$ 
	& High & Large \\
C & \ref{CR_0.19_c_0.0_Sep} & dotted & 30 & 15.2 & $128^2 \times 256$ 
	& Low  & Large \\
C & \ref{CR_0.19_c_0.0_Sep} & dashed & 40 & 11.5 & $128^2 \times 256$ 
	& High & Small \\
C & \ref{CR_0.19_c_0.0_Sep} & solid  & 40 & 15.5 & $172^2 \times 344$ 
	& High & Large \\
\end{tabular}
\end{center}
\caption{Grid sizes and resolutions. The resolution (Res.) is given
in number of grid points across the stellar diameter. The bounding box
length $B$ gives the extent of the physical space covered in each
direction in units of total rest mass $M_0$ (i.e.; the numerical grid
spans from$[~-B,~0.0,~0.0]$ to $[~B,~B,~B]$ since we make use of the
equatorial and $\pi$ symmetries of the systems.)}
\label{Table_Grids}
\end{table}
The ISCO parameters for each of the three cases are estimated as the
average of the parameters corresponding to the bracketing orbits
labeled {\it Stable} and {\it Merger} on each figure, while the
``errors'' span the difference.  We note that these ``errors'' are
partly due to numerical errors (see below), and partly by the
conceptional difficulty of defining a sharp ``ISCO''.  The results are
included in Table \ref{Table_InData}.  For the corotating sequence A
we also compare with the QE-ISCO at the QE turning-point.
The irrotational sequences considered in this paper terminate
before reaching a turning point (compare \cite{USE00,GGTMB01}).  The
termination of a sequence indicates that equilibrium models do
not exist at smaller separations, 
but since the numerical determination of this point relies on the
breakdown of a numerical (equilibrium) code, its accuracy is 
somewhat uncertain.
For both sequences we find that the ISCO is reached before the
sequence terminates.  For the irrotational sequence C one can also
compare with the first-order post-Newtonian ellipsoidal results of
Lombardi {\it et al.}  \cite{Lombardi:1997aw}, who find an angular
velocity of $\Omega m_0=0.0226$ for $(m/R)_{\infty} = 0.2$ binaries.

\begin{table*}
\begin{center}
\begin{tabular}{cccccccccc}
Case & rotation      & $m_0$ 	& $(m/R)_{\infty}$ & $M_i$ & $J_i/M_i^2$
	& $d/M_0$~~ & $\Omega ~m_0$ & $\Omega ~m_0$ (QE)  & $~~~f_{GW}~m_{1.4} ~(kHz)$\\
\tableline	
A    & corotating    & 0.1469	& 0.142 & $0.2708 \pm 0.0001$ & $1.08 \pm 0.01$
	& $9.0 \pm 0.8$~~ & $0.0162 \pm 0.0021$ & 0.0179 & 0.697 \\
B    & irrotational  & 0.1469	& 0.142 & $0.2702 \pm 0.0003$ & $0.97 \pm 0.04$
	& $9.4 \pm 0.9$~~ & $0.0154 \pm 0.0022$ & ---  & 0.662\\
C    & irrotational  & 0.1767	& 0.195 & $0.3171 \pm 0.0009$ & $0.93 \pm 0.04$
	& $7.7 \pm 0.8$~~ & $0.0199 \pm 0.0029$ & ---  & 0.838\\
\end{tabular}
\end{center}
\caption{Summary of our binary cases A, B, and C and results
for their ISCO.  For each sequence, we show the rotation state, the
rest mass of the individual stars $m_0$, the compaction in
isolation $(m/R)_{\infty}$, the total initial ADM mass $M_i$ and
angular momentum $J_i/M_i^2$, as well as the binary coordinate
separation at the ISCO $d/M_0$ (where $M_0$ is the total rest mass $2
m_0$) and the corresponding orbital angular velocity $\Omega ~m_0$.
For the corotational sequence we also compare with the QE
result for $\Omega ~m_0$ \cite{Marronetti:2003gk}. The wave frequency at the 
ISCO is $f_{GW}$ and $m_{1.4}$ is the stellar gravitational mass in
units of 1.4 $M_\odot$.}
\label{Table_InData}
\end{table*}

We consistently find that orbits become unstable before the onset of
instability as determined by QE methods, meaning at a smaller orbital
frequency.  This is not surprising, since the orbital decay becomes
fairly rapid just outside the QE-ISCO (compare, e.g., Fig.~2 in
\cite{DBSSU01}), so that our criterion for merger orbits applies to those
binaries.  This result is also consistent with earlier suggestions that the
transition through the ISCO may be fairly gradual \cite{gradual}.  The
similarity between the corotating and irrotational values for
the $(m/R)_{\infty} = 0.142$ orbits suggests that the dependence 
of the ISCO parameters on the stellar spins is not strong.

\begin{figure}
\epsfxsize=3.0in
\begin{center}
\leavevmode \epsffile{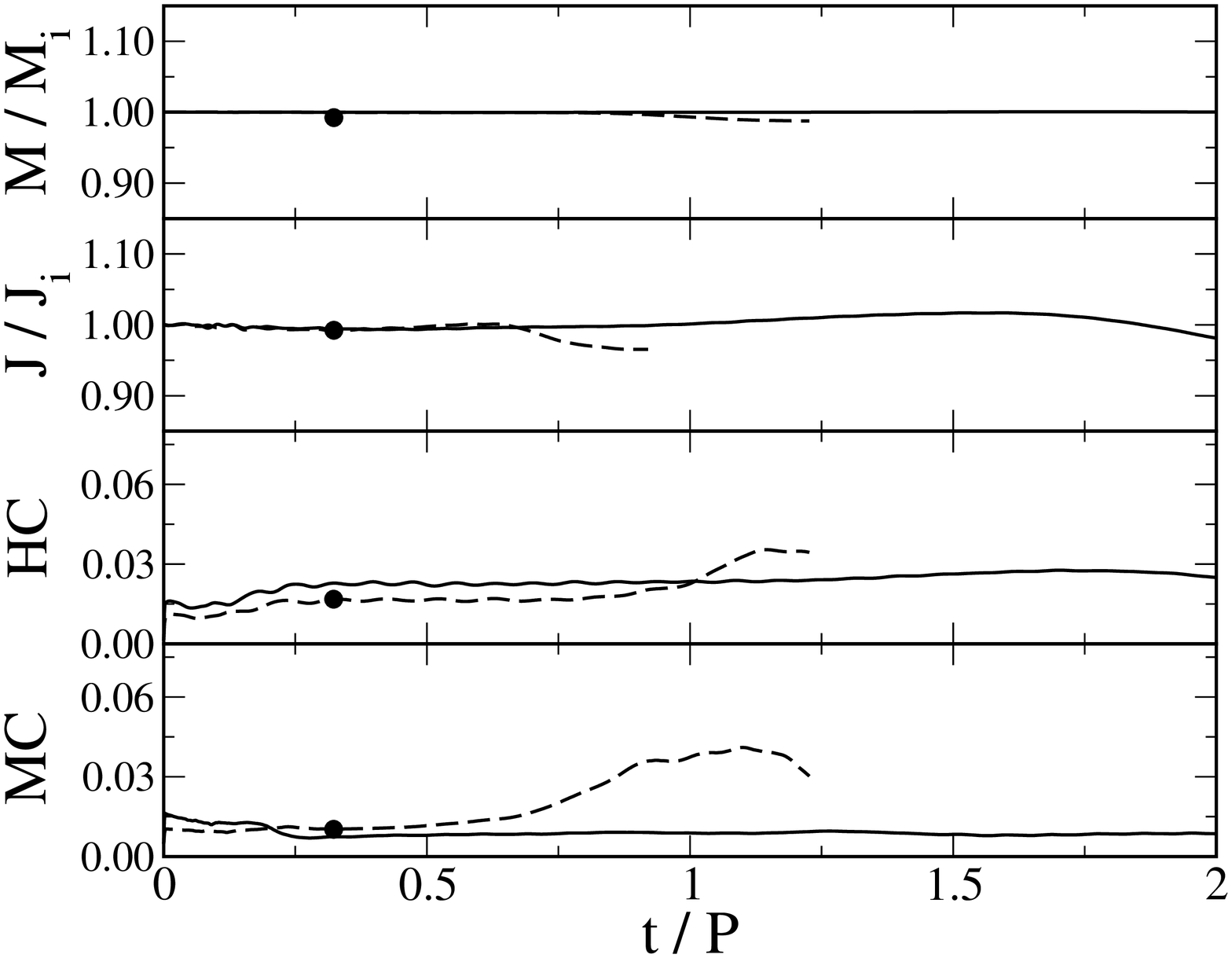}
\end{center}
\caption{ Quality control for the $(m/R)_\infty = 0.142$ irrotational
runs. We show from top to bottom the total gravitational mass, angular
momentum, the Hamiltonian constraint, and the average of the
components of the momentum constraint vs. time \cite{footnote4}. The
curves correspond to the runs from Fig. \ref{CR_0.14_a_1.0_Sep}
labeled as {\it Stable} (solid) and {\it Merger} (dashed). The filled
circle marks the time of surface contact for the merger orbit.}
\label{CR_0.14_c_0.0_QC}
\end{figure}

During all simulations we monitored the Hamiltonian and
Momentum constraints as well as the conservation of the total ADM mass
$M$ and angular momentum $J$ \cite{footnote5} (the rest mass $M_0$ is
conserved identically in our evolution scheme).  An example of these
for case B is shown in Fig.~\ref{CR_0.14_c_0.0_QC}.  In all our
simulations all quantities are conserved very well up to merger, after
which hydrodynamical effects including shocks and shear are  
handled only crudely by our artificial viscosity scheme.  
Stable runs ultimately
break down due to accumulation of numerical error.  We find that the
latter is sometimes dominated by hydrodynamical effects, leading to
deviations in the angular momentum, and sometimes by gravitational
effects, leading to violations of the constraint equations.
These effects are improved by increasing the grid resolution and
the separation to the outer boundaries, as well as using a coordinate
system that rotates with the binary as closely as possible.


{\bf Summary~} We present prototype simulations to determine dynamically the ISCO of
BNSs.  Evolving QE initial data at different separations, we bracket
the ISCO by distinguishing stable orbits, that remain in approximately
circular orbit for well over a period, from unstable ones, which decay
within a period.
The uncertainty in our results is caused both by numerical error and
the conceptual difficulty in defining a sharp ISCO.  Consistent with
earlier results \cite{gradual,DBSSU01} we find that binary orbits
start to plunge somewhat outside of the ``QE-ISCO'' as determined by
turning-point methods applied to QE initial data, resulting in a
cut-off in gravitational ``chirp'' signals at somewhat smaller
frequency.  Our preliminary results also seem to indicate that the
dependence of the ISCO parameters on the stellar spins is not
very strong.

One source of error is our assumption of
zero radial velocity in our binary initial data.  More realistic
initial data at finite binary separation would incorporate a radial
velocity that corresponds to the non-zero rate of inspiral at that
separation.  Miller \cite{Miller:2003pd} indicates that
this approximation may lead to non-negligible error, especially for
black hole binaries.  However, for the neutron star binaries and separations 
we consider here, the radial velocities would be at most $1-3\%$ of the
tangential velocity \cite{Miller:2003pd,DBSSU01}. 
Consequently, the error introduced by neglecting this component is 
likely to be smaller than the error bars already provided in Table \ref{Table_InData}.

It was recently pointed out that in dynamical evolutions of CTS
initial data describing corotating BNSs, the four-velocity $u^{\alpha}$
quickly deviates from being proportional to an exact helicoidal 
Killing vector, as assumed in constructing 
the initial models \cite{Miller:2003vw}. We confirm this result but find 
that this deviation arises from a small readjustment of the gravitational 
fields \cite{footnote6}, and not from appreciable changes in the density and velocity
profiles. We find no evidence of a significant breakdown of
quasiequilibrium for stable orbits and any spurious orbital 
eccentricities sharply decrease with increasing grid size.

\acknowledgments

It is a pleasure to thank Hwei-Jang Yo for useful discussions.  Most
of the calculations were performed at the National Center for
Supercomputing Applications at the University of Illinois at
Urbana-Champaign (UIUC).  This paper was supported in part by NSF
Grants PHY-0090310 and PHY-0205155 and NASA Grant NAG 5-10781 at UIUC
and NSF Grant PHY-0139907 at Bowdoin College. PM gratefully
acknowledges financial support through the Fortner Fellowship at the
University of Illinois at Urbana-Champaign (UIUC).


\begin{thebibliography}{99}
\bibitem{footnote2} We set $G=c=\kappa=1$, where $\kappa$ is the
polytropic constant of the EOS: $P=\kappa \rho_0^\Gamma$. For scaling
with $\kappa$, see ref \cite{BCSST98}.
\bibitem{footnote3} The time coordinate has been transformed into
fractions of the initial orbital period. Note that the periods differ
for binaries at different separation.
\bibitem{footnote5} The total angular momentum like the gravitational
mass is not strictly conserved due to the emission of
gravitational waves. However, the loss rate of angular momentum for a
typical orbit at these separations is of the order of $1\%$ to $2\%$.
Deviations exceeding these values therefore indicate numerical error.
\bibitem{footnote4} This figure presents the $L_2$ norm of the
constraint residuals. The details of these plots as well as the
normalizations used here can be found in
\cite{Duez:2002bn}. In this paper the momentum constraint curve shows
the average of the three spatial components.
\bibitem{footnote6} E.g., the loss of a small amount of spurious radiation 
present in the initial data, which leaves M and J nearly unchanged.
\end{thebibliography}
\end{document}